\documentclass[11pt]{article}
\usepackage[top=1in, bottom=1in, left=1in, right=1in]{geometry}                % See geometry.pdf to learn the layout options. There are lots.
\pdfoutput=1
\usepackage{graphicx}
\usepackage{amsmath}
\usepackage{amssymb}

\usepackage{fancyheadings}

\newcommand{\ket}[1]{\left|#1\right\rangle}

\title{Emergent Fractional Charge and Multiple Majoranas}
\author{R. Jackiw\\
\it \small Center for Theoretical Physics\\
\it \small Massachusetts Institute of Technology\\
\it \small Cambridge, MA 02139}
\date{}                                           % Activate to display a given date or no date

\begin{document}
\maketitle
\thispagestyle{fancy}

\begin{abstract}
To mark the 111$^{th}$ birthday of Eugene Wigner, we review topological excitations in diverse dimensions.
\end{abstract}

Although this meeting commemorates the 111 birthday of Wigner, my talk draws on the work of Dirac. I do not  apologize for this. Both men (brothers-in-law) were led by mathematics to descriptions of fundamental particles: Wigner by group theory, Dirac by his differential equation. It is Dirac's framework that recently resulted in further discoveries about  quantum excitations. This is the topic that I shall describe. 

Let me begin with a bit of history about Dirac and his equation. Dirac was seeking a relativistic equation for electrons and eventually arrived at the  following first-order matrix/differential equation.

\begin{changemargin}{-33pt}{33pt} 
\includegraphics[scale=.9]{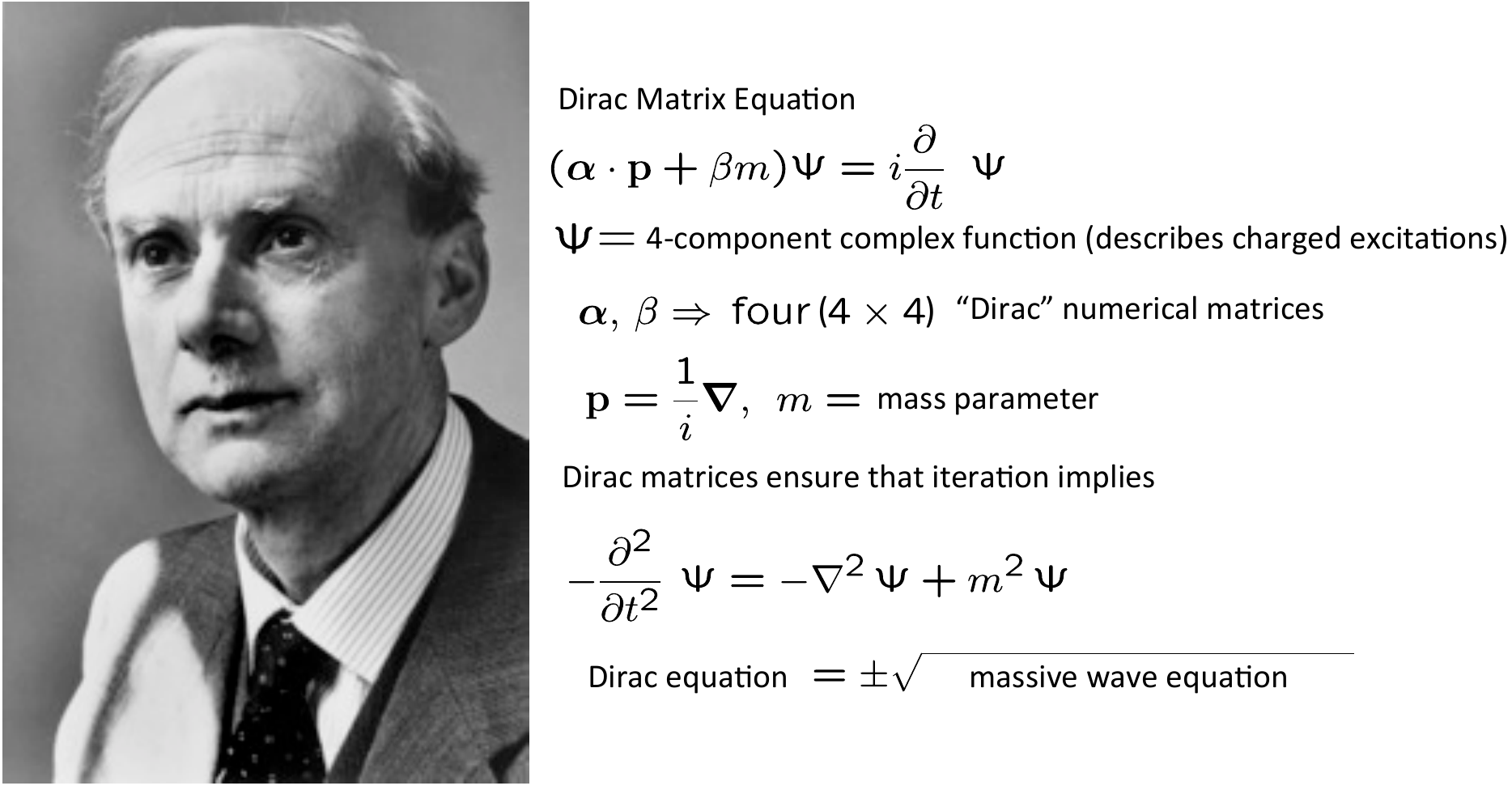}
\end{changemargin}

To expose properties of the Dirac equation, we make the usual decomposition
\[
\Psi = e^{-iEt}\, \psi  \Rightarrow [{\boldsymbol \alpha \cdot \mathbf{p}} + \beta\, m] \psi = E \psi
\]
and discover that there exist positive energy solutions $E>0$, which can describe electrons. 

\noindent But because ``square roots" come with both signs, there exist also negative energy solutions, $E<0$. These need interpretation because they cannot describe electrons, which carry positive energy. After some hesitation, Dirac concluded that the negative energy solutions correspond to anti-electrons, {\it viz.} positrons, which soon were discovered.  

In a further conceptual leap, Dirac posited that in the ground state all negative energy levels are filled, but nevertheless the charge of the ground state is zero. This means that the Dirac equation is really a many particle equation, where the particles populate the single particle energy levels.

The Dirac equation is a beautiful equation, with a rich hidden physical structure. It goes beyond a single particle interpretation and it predicts new (anti-)particles: the positrons. These characteristics make it a beautiful equation for physics.  Also it is a beautiful equation for mathematics because by using matrices, it succeeds in taking a square root of a second order differential equation. The mathematical and physical beauty of the Dirac equation suggests to mathematicians and physicists that deformations of the equation may also yield beautiful and interesting results. But which deformation should we consider?

 One alteration that can be made is to reconsider the equation in dimensions different  from the three spatial and one time dimension. If we take fewer dimensions we gain the mathematical advantage of simplicity and also have the possibility  of describing physical systems that are confined to lower dimensions, for example to a line or to a plane. Such configurations can occur in condensed matter physics. There one encounters situations where the low energy dynamics is well described by a matrix equation, linear in the momenta.  Depending on the nature of the material, the equation may describe excitations on a line, on the plane in addition to those in the three-dimensional bulk.

Thus the role of the Dirac equation expands from its original task of describing electrons ($E>0$) and positrons ($E<0$) in three dimensions, to a model for condensed matter, with bound valence-band electrons ($E<0$) and conduction-band electrons ($E>0$). Also there may be a constant mass term, which separates the positive energy solutions from the negative energy ones by a 2 $|m|$ ``gap."    The charge $Q$ of the empty vacuum vanishes.

A further more profound deformation allows the mass term $m$ to depend on position. Of course  a weak dependence will produce only insignificant changes from the usual homogenous mass case; we are interested in a significant dependence on  position, which could significantly alter the physical situation. Note that the gap $2\, |m|$ depends only on the magnitude of $m$, and not on its sign. ${\scriptscriptstyle{+}}\, m$ produces the same gap as -$m$. This suggest a deformation of the mass term that interpolates between positive and negative values. In this way we are led to a Dirac equation in the presence of a topologically non-trivial defect.

\centerline{$m \to \varphi \, ({\bf r})$}
\vskip 1ex

\centerline{$[\boldsymbol{\alpha}\cdot \mathbf{p} + \beta\, \varphi\, (\mathbf{r})]  \psi = E \psi$}

\vskip 4ex
\includegraphics[scale=.50]{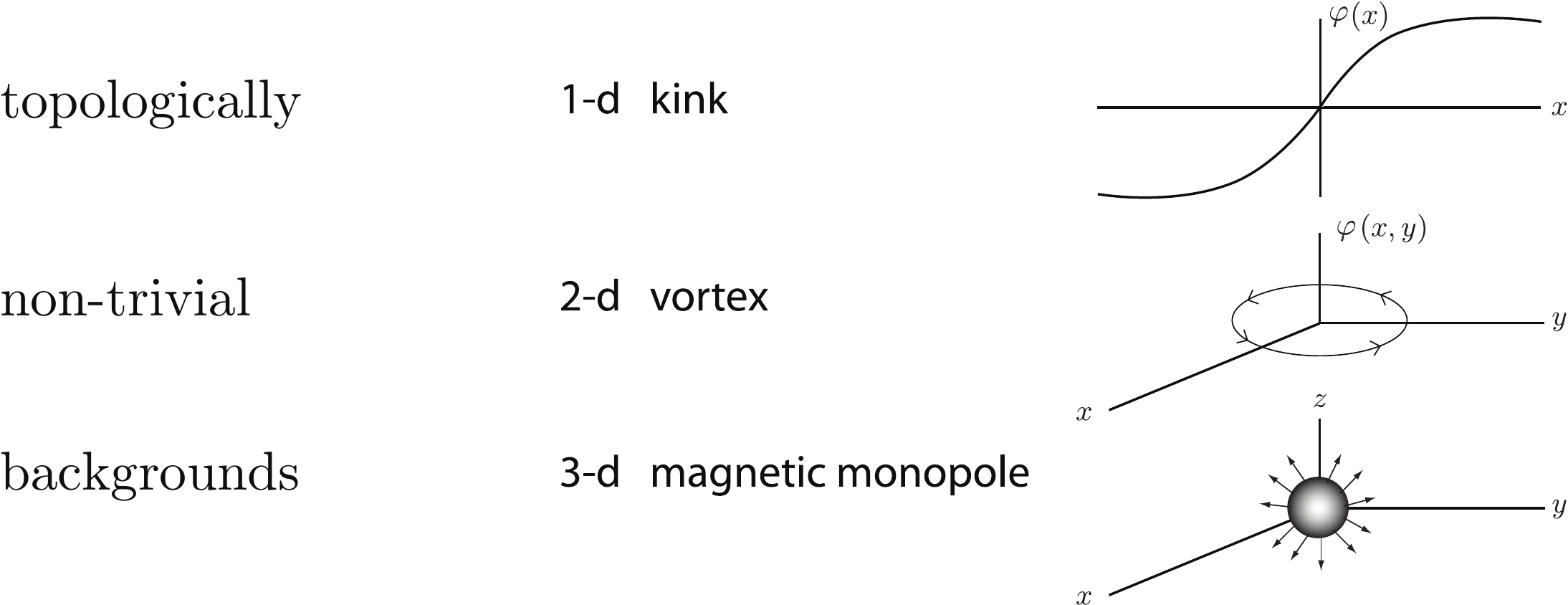}\\

To determine the nature of the excitations that are described by the above  Dirac-like equation in the presence of topologically non-trivial background, we study the energy eigenvalues and find as before continuum solutions with $E>0$ and $E<0$. But additionally there exists one (or more) normalizable, isolated $E=0$ mode. This mid-gap state can be found by explicit solution to the differential equation. But remarkably the result can also be established by mathematical ``index" theorems, which count the number of zero-energy modes, in terms of topological properties of the background $\varphi (\mathbf{r})$.

The presence of the zero-energy state raises a central question: in the vacuum is the mid-gap state empty or filled; {\it i.e.} does the state belong to the positive $ E$ solutions, or to the negative $ E$ solutions? Also what is the charge? Dirac has nothing to say about this -- he did not know about zero-energy modes in his equation.

The unexpected answer was given by Rebbi and me, [{\it PRD} {\bf 13}, 3398 (76)]. The empty state carries charge $ Q= -\frac{1}{2}$, when it is filled $Q = + \frac{1}{2}$.

There are many ways to prove this result. They range from bond-counting in the presence of defects to a fully quantum field theoretic argument. I shall not  present any of these proofs here. But it is important to state that the fractional charge is an eigenvalue, not merely an expectation value.

The discovery of charge fractionalization opened physicists' minds to the possibility, and actual experimental investigation has been proposed. Firstly in one dimension, Su, Schrieffer \& Heeger [{\it PRL} {\bf 42}, 169 (79)] made use of this mechanism in their explanation of conductivity in polyacetylene. In two dimensions fractional charge has been suggested for graphene [Chamon, Hou \& Mudry, {\it PRL} {\bf 98} 186809 (07)] and  established for the quantum Hall effect. [Hall effect fractionalization does not rely on topological defects, but is inspired by them.] Three dimension still awaits experimental application.

 The original Dirac equation involves complex fields, which describe fermion particles  (electrons) and anti-particles (positrons). On the other hand, there are chargeless, neutral bosons that are their own anti-particles. Examples are the spin zero neutral pion, the spin one chargeless photon, and the hypothetical spin two chargeless graviton. In quantum field theory, all of these bosons are described by real fields. It is natural to inquire how we should describe neutral fermions that are their own anti-particles. Admittedly there is no direct experimental evidence for such fermions. But present-day theory makes use of them to account for recent developments in neutrino physics (non-vanishing mass, non-conservation of individual lepton number). Also supersymmetric partners of the self-conjugate bosons must be self-conjugate fermions.
 
 \newpage
\includegraphics[scale=.30]{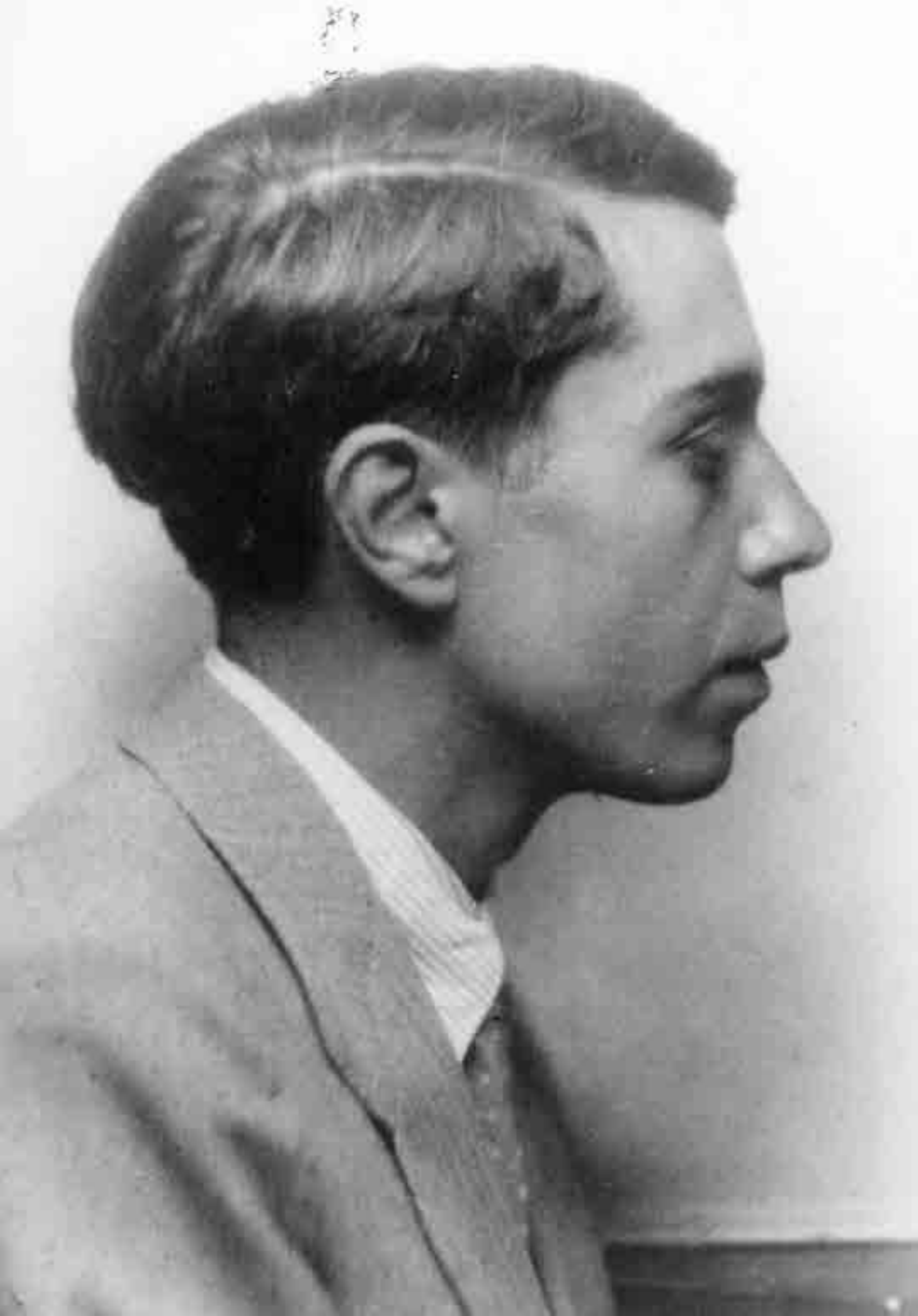} \ 
\begin{minipage}[b][1in][b]{3.75in}
%\singlespace
 The Majorana  equation gives a description of self-conjugate fermions in terms of a real Dirac equation. Recall that equation
 \[
 (\boldsymbol{\alpha}\cdot \mathbf{p} + \beta m) \ \Psi = i \frac{\partial}{\partial t}\ \Psi
 \]
$\Psi$ can be real if $\beta$ is imaginary, and since $\mathbf{p} =- i\,  \boldsymbol{\nabla},  \boldsymbol{\alpha}$ must be real. Such a choice of Dirac matrices defines the Majorana representation.

An explicit example of the Majorana representation for such matrices is 
\begin{eqnarray*}
\alpha^1_M = \left(\begin{array}{cc}0 & \sigma^1 \\ \sigma^1 & 0\end{array}\right)\ 
\alpha^2_M = \left(\begin{array}{cc}I & 0 \\ 0 & -I\end{array}\right)\
\alpha^3_M = \left(\begin{array}{cc}0 & \sigma^3 \\ \sigma^3 & 0\end{array}\right)\\[1ex]
\beta_M = \left(\begin{array}{cc}0 & \sigma^2 \\ \sigma^2 & 0\end{array}\right)\quad \Psi_M^\ast = \Psi_M \hspace{3em}
\end{eqnarray*}
\[
{\boldsymbol\alpha}_M = {\boldsymbol\alpha}_M , \beta_M = - \beta_M
\]
\end{minipage}
\bigskip

\noindent However, since the physical content is invariant against similarity  transformations, the above reality requirements can be replaced by a reality up to similarity, effected by a conjugation matrix $C$.
\[
C \boldsymbol{\alpha}^\ast \, C^{-1} = \boldsymbol{\alpha}, \ C \beta^\ast \, C^{-1}  = -\beta \quad C \Psi^\ast = \Psi
\]
For example, if we use the Weyl representation
\[
\boldsymbol{\alpha} = \left(
                                                                   \begin{array}{cc}
                                                                        \boldsymbol{\sigma} & 0 \\  
                                                                        0 & -\boldsymbol{\sigma}
                                                                        \end{array}\right)\ , \
\beta = \left(
                 \begin{array}{cc}
                         0& I \\ 
                         I & 0
                         \end{array}\right)
\]

\noindent the conjugation matrix is  $C= \left(\begin{array}{cc}0 &- i\sigma^2 \\ i\sigma^2 & 0\end{array}\right)$. (In the Majorana representation $C = I$.)

Thus the Majorana equation retains its Dirac/Weyl form, except that the condition $C \Psi^\ast =\Psi$ is imposed.
\begin{eqnarray*}
%C \, \psi^\ast = \psi \Rightarrow \hspace{3.5in}\\
\Psi = \left(\begin{array}{c}\psi \\ \chi\end{array}\right)\qquad\
\left(
                 \begin{array}{cc}
                         \boldsymbol{\sigma} \cdot \boldsymbol{p}& m  \\ [1ex]
                         m & - \boldsymbol{\sigma} \cdot \boldsymbol{p}
                         \end{array}\right)
                         \left(\begin{array}{c}\psi \\ \chi\end{array}\right) =
                         i \frac{\partial}{\partial t}\ \left(\begin{array}{c}\psi \\ \chi\end{array}\right)\\[1ex]
                         C \, \Psi^\ast = \Psi \Rightarrow \chi = i\, \sigma^2\, \psi^\ast \hspace{2.10in}
                         \end{eqnarray*}
                         \text{This leads to the two-component}\\                         
                         \text{\hspace{5em}\textsf{Majorana Matrix Equation}.}%\\[1ex]
                         \begin{eqnarray*}
                         \boldsymbol{\sigma}  \cdot \boldsymbol{p} \, \psi + i \sigma^2\, m \, \psi^\ast =  i \frac{\partial}{\partial t}\ \psi\qquad (2\times 2)
\end{eqnarray*}
Note that $\psi$ mixes with $\psi^\ast$; the Majorana mass term does not preserve any quantum numbers; there is no distribution between particles and anti-particle since there are no quantum numbers to tell them apart; {\it i.e.} particle is its own anti-particle.

The field expansion for a charged Dirac field reads
 \[
   \Psi = \sum_{E>0} \left(a_E\, e^{-i Et}\, \Psi_E + b^\dagger_E\, e^{i Et}\, C\, \Psi^\ast_E\right) ,
    \]
with $a$ annihilating and $b^\dagger$ creating particles and anti-particles respectively. For the Majorana field we have
 \[
    \Psi = \sum_{E>0} \left(a_E\, e^{-i Et}\, \Psi_E + a^\dagger_E \, e^{i Et}\, C\, \Psi^\ast_E\right)  .
    \]
The anti-particle operators ($b, b^\dagger$) have disappeared!

The remarkable fact is that the condensed matter theorists have encountered essentially the same equation in a description of a super conductor in contact with a topological insulator. The relevant two dimensional Hamiltonian density reads
\[
H = \psi^{\ast}\ ({\boldsymbol\sigma} \cdot \frac{1}{i}\ {\boldsymbol\nabla} -\mu)\ \psi + \frac{1}{2}\ (\triangle \psi^{\ast}\, i\, \sigma^2 \, \psi^\ast + h.c.) .
\]
 $\psi=
\left(\begin{array}{c}
\psi_{\scriptscriptstyle \uparrow}\\ \psi_{\scriptscriptstyle \downarrow}
\end{array}\right) , 
{\boldsymbol\sigma} = (\sigma^1, \sigma^2), \mu $ is chemical potential and $\triangle$ is the order parameter that may be constant: $\triangle = \triangle_0$, or takes vortex profile; $\triangle ({\bf r}) = v (r) e^{i\theta}, v(0) = 0, v (\infty) = \triangle_0$.

The equation of motion follows: 
\[
\ i\, \partial_t \,  \psi = ({\boldsymbol \sigma} \cdot {\boldsymbol {\bf p}} -\mu)\ \psi + \triangle\, i\, \sigma^2 \, \psi^\ast  
\]

\noindent In the absence of $\mu$, and with constant $\triangle$ , the above system is a (2+1)-dimensional version of the  (3+1)-dimensional, two component Majorana equation! It governs chargeless spin $\tfrac{1}{2}$ fermions with  Majorana mass $|\triangle|$.

In the presence of a single vortex order parameter $\Delta ({\bf r}) = v (r) e^{i\theta}$ there exists a zero-energy (static) isolated mode
 \big[Rossi \& RJ {\it NPB} {\bf 190}, 681 (81); Fu \& Kane, {\it PRL} {\bf 100}, 096407 (08)\big] 
 \[
 \psi_0 = \# \left(
                  \begin{array}{l}
                  J_0 (\mu r) \exp\, \{-i\pi/4 - V(r)\}\\[1ex]
                  J_1 (\mu r) \exp\, \{i (\theta + \pi/4) - V(r)\}
                  \end{array}
                  \right)
 \]
 $\#$ real constant, $V^\prime (r) = v (r)$. The  Majorana field expansion now reads:\\
 %\vspace{1ex}
 % \hspace{2in}  \Psi = ................ \quad + a \, \psi_0\\
% $

\[
 \begin{array}{ccccc}
\Psi & =\!\! &\negthickspace\!\! ................ \negthickspace\!\! &  + \ a \, \psi_0 & \\
&& {{\scriptstyle E \ne 0} \ \text{\small modes}}&& 
\end{array}
\]

%$
% \vspace{-2ex}
% \hspace{4in}${\scriptstyle E \ne 0}$ {\scriptsize modes}\\
 where the zero mode operator $a$ satisfies
 \[
 \{a, a^\dagger\} = 1, a^\dagger = a \Rightarrow a^2 = 1/2 .
 \]
 How to realize $a$ on states? 
 There are two possibilities [Chamon, Nishida, Pi, Santos \& RJ; {\it PRB} {\bf 81}, 224515 (10)].

\begin{itemize}
\item[(i)] Two 1-dimensional realizations: take vacuum state to be eigenstate of $a$, with possible eigenvalues $\pm 1/\sqrt{2}$.
\[
a\ket{0\pm} = \pm\ \frac{1}{\sqrt{2}}\ \ket{0\pm}
\]
There are two ground states $\ket{0+}$ and $\ket{0-}$. Two towers of states are constructed by repeated application of $a^\dagger_E$. No operator connects the two towers. 
Fermion parity is broken because $a$ is a fermionic operator. Like in spontaneous breaking, a   vacuum $\ket{0+}$ or $\ket{0-}$ must be chosen, and no tunneling connects to the other ground state.
\item[(ii)] One 2-dimensional realization: vacuum doubly degenerate $\ket{1}, \ket{2}$, and $a$ connects the two vacua.
\[
\begin{array}{l}
a \ket{1} = \frac{1}{\sqrt{2}}\ \ket{2}\\[2ex]
a \ket{2} = \frac{1}{\sqrt{2}}\ \ket{1}
\end{array}
\]
Two towers of states are constructed by repeated application of $a^\dagger_E$. $a$ connects the towers. Fermion parity is preserved.
\end{itemize}
It seems natural to assume that fermion parity is preserved. Therefore, we adopt the second possibility, which may also be justified by considering a vortex/anti-vortex pair separated by a large distance. 

But we note that fermion parity violating realization (i) has a place in mathematical physics: Let $\mathcal{L}$ be Lagrange density for scalar kink $\oplus$ fermions.
\begin{itemize}
\item[]
%\vspace{-1.5ex}
	\begin{itemize}
		\item[]  %scalar kink $\oplus$ fermions\\[1ex]
		$\mathcal{L} = \frac{1}{2}\ \partial_\mu\, \Phi\, \partial^\mu\, \Phi + \frac{\mu}{2}^2\, \Phi^2 - \frac{\lambda}{8}^2\, \Phi^4 + i \bar{\Psi}\, \gamma^\mu\, \partial_\mu\, \Psi - g\, \Phi\, \bar{\Psi} \Psi$%\bigskip
		\item[] $\mathcal{L}$ possesses SUSY for $g = \lambda, \Psi$ Majorana
		\end{itemize}		
	\end{itemize}
 One can prove from center anomaly in SUSY algebra that fermion parity can be absent [Losev, Shifman \& Vainshtein, {\it PLB} {\bf 522}, 327 (01)].\negthinspace\ It is an open question whether this curiosity has any relevance for condensed matter, indeed for any physical question. \big(For some speculation see [Semenoff \& Sodano, {\it EJTP} {\bf 10}, 57 (08)].\big)

How many states $\mathcal{N}$ are needed to represent $N$ vortices (in a fermion parity preserving fation)? For $N=1$ we used two states: $\mathcal{N}=2$; for $N=2$, we have Hermitian $a_1$ and $a_2$, with $a^2_1 = a^2_2 = \frac{1}{2}$ and $a_1\, a_2 + a_2 \, a_1 = 0$. These two can be realized on the two states that are already present at $N=1$.
%\vspace{2ex}
\begin{center}
\begin{tabular}{rrrl}
&&& $a_1 \ket{1} = \frac{1}{\sqrt{2}} \ket{2} \  \ a_1 \ket{2} = \frac{1}{\sqrt{2}} \ket{1}$\\[1.5ex]
&&&$ a_2 \ket{1} = \frac{i}{\sqrt{2}} \ket{2} \  \ a_2 \ket{2} =  \frac{-i}{\sqrt{2}} \ket{1}$
\end{tabular}
\end{center}

We can present these formulas explicitly by denoting the states by Cartesian 2-vectors, and the operators $a_i$ by Pauli matrices.
\begin{center}
\begin{tabular}{rrrl}
&&&$\ket{1} \sim\left(
\begin{array}{c}
 1 \\[-1ex]
  0
\end{array}\right) \ \ \ket{2} \sim \left(
\begin{array}{c}
 0 \\[-1ex]
  1
\end{array}\right)$\\[3.25ex]
&&& $a_1 =  \frac{\sigma^1}{\sqrt{2}} \  \ a_2 =  \frac{\sigma^2}{\sqrt{2}}$
\end{tabular}
\end{center}
These results verify the formulas  $\mathcal{N} = e^{\frac{N}{2}}$ for $N$ even and $\mathcal{N}= e^{\frac{N+1}{2}} $ for $N$ odd.

For $N=3$, we have three mode operators: $a_i, i= 1, 2, 3$. We cannot use three Pauli matrices to represent them; in particular we cannot set $a_3 = \frac{\sigma^3}{\sqrt{2}}$ because $\sigma^3$ is diagonal on the above Cartesian states and would lead to fermion parity violation. So for $N=3$ we must use $4 \times 4$ Dirac matrices and Cartesian 4-vectors.
\[
\text{states:} \ \ \ket{1} \sim
\left(
\begin{array}{c}
 1 \\[-1ex]
  0\\[-1ex]
  0\\[-1ex]
0
\end{array}
\right) \, ,
\ket{2} \sim
\left(
\begin{array}{c}
 0 \\[-1ex]
  1\\[-1ex]
  0\\[-1ex]
0
\end{array}
\right) \, , 
\ket{3} \sim
\left(
\begin{array}{c}
 0 \\[-1ex]
  0\\[-1ex]
  1\\[-1ex]
0
\end{array}
\right) \, ,
\ket{4} \sim
\left(
\begin{array}{c}
 0 \\[-1ex]
  0\\[-1ex]
  0\\[-1ex]
1
\end{array}
\right)
\]
\[
\text{operators:} \ \ \boldsymbol{\alpha}  =\left(
						\begin{array}{cc}
						0  & i {\boldsymbol \sigma}  \\
						- i {\boldsymbol \sigma}  & 0 
						\end{array}
						\right), \ \ \beta =  \left(
						\begin{array}{cc}
						 0 & I \\
						 I & 0
						\end{array}
						\right)\hspace{10em}
\] 
Choose 3 of 4 Dirac matrices $\frac{1}{\sqrt{2}} ({\boldsymbol\alpha}, \beta)$. They square to $\frac{1}{2}$, anti-commute with each other, and act on the 4 basis vectors. Thus we 
verify $\mathcal{N} = 2^{\frac{N+1}{2}} = 4 \ \text{for} \ N= 3$.

 \vtop{\hsize=6in [The matrix $\left(
                                                                               \begin{array}{cc}
							 I  &0  \\
							0 & -I 
							\end{array}
							\right)$
	also anti-commutes but cannot be used because it is diagonal, and would lead to fermion parity violation.]}

\vspace{2ex}	
 Higher N:  The pattern is now clear. We use a Clifford algebra realized by $\mathcal{N} \times \mathcal{N}$ matrices and $\mathcal{N}$-component  Cartesian vectors to represent $N$ zero-mode operators acting on states.
In selecting the members of the Clifford algebra, diagonal elements (in the Cartesian basis) must not be used, because they correspond to fermion parity violating realizations.
The fermion parity preserving realization leads to a number of states given by formulas $\mathcal{N} = 2^{\frac{N}{2}}$ and $\mathcal{N} = 2^{\frac{N+1}{2}}$ for even number and odd number vortices, respectively. The emergent algebra is a Clifford algebra, with the restriction that diagonal elements are not present [Pi \& RJ, {\it PRB} {\bf 85}, 033102 (12)].

It is an interesting open question, what role, if any, should be assigned to the fermion parity violating realizations, which arise in supersymmetry. 

These days we anticipate hearing from experimentalists that Majoranas have been found. But who will be first: condensed matter or particle physicists?
\end{document}